\documentclass[12pt,preprint]{aastex}

\shorttitle{LGS GLAO at the MMT}
\shortauthors{Baranec et al.}

\begin{document}


\title{On-sky wide field adaptive optics correction using multiple laser guide stars at the MMT}


\author{Christoph Baranec\altaffilmark{1,2}, Michael Hart\altaffilmark{1}, N. Mark Milton\altaffilmark{1}, Thomas Stalcup\altaffilmark{3},
        Keith Powell\altaffilmark{1}, Miguel Snyder\altaffilmark{1,4}, Vidhya Vaitheeswaran\altaffilmark{1}, Don McCarthy\altaffilmark{1} and
        Craig Kulesa\altaffilmark{1}}
\email{baranec@astro.caltech.edu}


\altaffiltext{1}{Steward Observatory, University of Arizona, Tucson, AZ 85721}
\altaffiltext{2}{now at Caltech Optical Observatories, California Institute of Technology, Pasadena, CA 91125}
\altaffiltext{3}{MMT Observatory, Tucson, AZ 85721}
\altaffiltext{4}{now at the Department of the Army}


\begin{abstract}
We describe results from the first astronomical adaptive optics system to use multiple laser guide stars, located at the 6.5-m MMT telescope in Arizona.
Its initial operational mode, ground-layer adaptive optics (GLAO), provides uniform stellar wavefront correction within the 2 arc minute diameter laser 
beacon constellation, reducing the stellar image widths by as much as 53\%, from 0.70 to 0.33 arc seconds at $\lambda = 2.14$ $\mu$m. GLAO is achieved by 
applying a correction to the telescope's adaptive secondary mirror that is an average of wavefront measurements from five laser beacons supplemented 
with image motion from a faint stellar source. Optimization of the adaptive optics system in subsequent commissioning runs will further improve 
correction performance where it is predicted to deliver 0.1 to 0.2 arc second resolution in the near-infrared during a majority of seeing conditions.
\end{abstract}





\section{Introduction}

Until very recently, adaptive optics (AO) systems supporting astronomical observing have used single guide stars to measure atmospheric turbulence.
This has limited the best optical correction to a single target, with correction degrading as a function of increasing field angle. By using multiple guide
stars, many types of wide field adaptive optics correction can be implemented. The simplest of these, ground-layer adaptive optics (GLAO), was suggested
by Rigaut \cite{rigaut} as a way to improve wide field imaging for large telescopes. Wavefront measurements from guide stars located far from each
other (2 -- 10+ arc minutes) can be averaged to estimate the turbulence close to the telescope aperture. A partially corrected field over the guide star
constellation is produced when only this low-lying turbulence is compensated. Alternatively, single Rayleigh laser guide star (LGS) systems at smaller
telescopes such as the 4.2-m William Herschel Telescope\cite{WHT} and planned for the 4.1-m Southern Astrophysical Research\cite{SAM}
telescope can produce wavefront compensation over fields of 1.5 and 3 arc minutes respectively. However, a single beacon GLAO implementation
will not be effective for larger apertures which suffer from stronger focal anisoplanatism effects that limit the accuracy of the ground-layer
turbulence measurements\cite{andersen}. Multiple laser guide stars will also be required for GLAO implementations that achieve full sky coverage 
on the next generation of extremely large (25+ m) class telescopes.

Simulations have predicted that multiple-beacon GLAO will effectively and consistently improve the atmospheric seeing
\citep{rigaut, andersen, louarn, tokovinin} and open-loop studies at the 6.5-m MMT and 1.5-m Kuiper telescopes predict that GLAO can reduce
wavefront aberration by up to $\sim 45\%$ \cite{apj, spie, ngsglao}. Using simultaneous wavefront measurements of 5 laser guide stars and a stellar
source at the MMT, synthetic point-spread-functions (PSFs) have been calculated from the residual wavefront errors of open-loop ground layer correction
\cite{opex, spie}. During roughly median seeing conditions ($r_0 = 14.8$ cm at $\lambda = 500$ nm), the full-width at half-maximum (FWHM) of the
corrected PSF was calculated to be 0.12 and 0.16 arc seconds in the $K$ and $H$ bands respectively. Azimuthally averaged PSF profiles of both
the uncorrected and open-loop ground-layer corrected stellar PSFs are presented in figure 1. During more favorable seeing conditions 
($r_0 = 22.6$ cm), the calculated FWHM size of the open-loop ground-layer corrected PSFs were 0.10 and 0.14 arc seconds in the $K$ and $H$ bands.
Although the performance of GLAO is affected by both the changing seeing conditions and the relative strength of ground-layer turbulence, it has been found
at many sites, including at the nearby Mt. Graham and Mt. Bigelow, that typically one-half to two-thirds of the atmospheric turbulence lies in this ground layer
\cite{andersen, avila, egner, tok_cn2a, tok_cn2b, ngsglao, velur}. Further statistics on the strength of the ground-layer at the MMT need to be measured; however
it is expected that the GLAO system will be capable of delivering 0.1 to 0.2 arc second resolution in the near-infrared during a majority of seeing conditions.

GLAO was first demonstrated on-sky in closed-loop using three bright natural guide stars on a 1.5 arc minute diameter
with the Multi-conjugate Adaptive Optics Demonstrator fielded at the Very Large Telescope (VLT) in early 2007 \citep{mad1,mad2}. Strehl ratios 
from 5 to 22.5$\%$ in $K$ band were observed within the guide star constellation during seeing of $\sim$ 0.7 arc seconds. However, 
sky coverage and the number of accessible science targets will be limited because there are very few suitably bright natural guide star constellations.

The MMT multiple laser adaptive optics system is being commissioned to support wide-field science with GLAO and narrow-field science with laser
tomography adaptive optics (where Strehl is maximized at the cost of corrected field-of-view). In February 2008, the system demonstrated an initial
GLAO correction where stellar image widths were reduced by as much as 53\%, from 0.70 to 0.33 arc seconds at $\lambda$ = 2.14 $\mu$m. Further
commissioning in May 2008 demonstrated wide field aberration compensation of a constellation of stars, and a reduction of the stellar wavefront 
errors in the control space of the laser wavefront sensor by $38\%$. The observations reported here demonstrate significant image improvement across
a 110 arc second field, but not yet at the anticipated resolution of 0.1 to 0.2 arc seconds. At the time of these observations, the GLAO system was
affected by several issues in the tip-tilt loop which compromised the overall closed-loop performance. The primary problem were network delays and dropouts
between the tip-tilt sensor and real-time reconstructor, responsible for unwanted random control loop delays, causing the system to be unstable at anything other
than at very low control loop gain values. The tip-tilt issues have since been diagnosed and corrected; thus future observations should not be affected. 

\section{Experimental design}
\subsection{Instrument description}

The MMT's multiple laser guide star AO system comprises a laser launch telescope\citep{stalcup}, an adaptive secondary mirror (ASM)\citep{brusa,wildi},
a Cassegrain mounted wavefront sensor instrument \\\cite{opex,apj,spie,baranec}, and a PC based real-time reconstructor computer\cite{pcr}. The laser launch telescope,
mounted above the MMT's ASM, projects a constellation of five $\lambda = 532$ nm Rayleigh LGSs on a regular 2 arc minute diameter pentagon with a
total on-sky power of 25 W.

The Cassegrain wavefront sensor instrument consists of a laser guide star wavefront sensor (LGS WFS), a tip-tilt sensor and a natural guide star wavefront 
sensor (NGS WFS). The LGS WFS uses dynamically refocused optics\cite{georges} and an electronically shuttered CCD to accumulate photon return over
a range of $20-29$ km from the telescope. Shack-Hartmann patterns for each of the five LGSs are created by a prism array\citep{prism} with $60$ 
subapertures in a hexapolar geometry, and captured at a rate of 400 frames per second. An electron multiplying CCD is used to obtain tip-tilt measurements
from a natural star within the $2$ arc minute LGS constellation at the same rate. The NGS WFS is a traditional $12 \times 12$ Shack-Hartmann sensor used
for system calibration and automatic static aberration correction. The Cassegrain mounted wavefront sensor instrument is designed to accept the current suite
of MMT $f/15$ NGS AO science instruments including PISCES\citep{pisces}, Clio\citep{clio}, ARIES\citep{aries} and BLINC-MIRAC\citep{blinc}.

The GLAO correction is calculated by reconstructor matrix multiplication in the PC based real-time computer from $300$ LGS slope pairs, $60$ pairs from
each of the five laser beacons, as well as a pair of slopes from the fast tip-tilt camera. The reconstructor projects each of the five sets of LGS WFS slope
measurements onto an orthonormal basis of disk harmonic (DH) functions\citep{DH} in the telescope's pupil. The five LGS wavefronts are averaged 
by mode to produce an estimate of the ground-layer contribution to atmospheric seeing. Finally, the GLAO modal estimate is converted to actuator 
displacements which are transmitted to the ASM at the telescope pupil. In addition to an overall system loop gain, separate gains can be applied to the 
individual DH modes in the reconstruction for fine tuning of the system response. These are obtained from the measured modal closed-loop system response.
Measured scale factors for the
individual responses of the ASM actuators are also applied to account for variations in the sensitivities of the associated capacitive sensors. The DH 
basis functions are used instead of the traditional Zernike polynomials since they provide increased loop stability and a lower RMS error for a given
number of controlled modes. Zernike polynomials have large radial derivatives near the edge of the pupil, particularly for high spatial frequency modes.
By contrast, the DH functions have zero radial derivatives at the edge of the pupil, placing less stress on actuators at the outer edge of the ASM and 
resulting in lower actuator currents and greater loop stability.

\subsection{MMT seeing measurements}

The results of GLAO correction are presented in the context of seeing statistics recorded at the telescope 
spanning the past 5 years. Measurements of seeing are made by calculating stellar image widths directly as observed by the Shack-Hartmann wavefront
sensors used for the active alignment of the MMT's primary mirror with its $f/5$ and $f/9$ secondary mirrors \cite{MMTseeing}. The sensors do not
have the temporal bandwidth to resolve the dynamic atmospheric turbulence so the integrated image widths represent the best available estimate
of seeing. The wavefront sensors currently operate at a mean wavelength of $\overline{\lambda} = 650$ nm (T. Pickering 2008, private communication)
and data comprise $\sim 50000$ individual measurements taken during approximately 800 different nights. Measurements are corrected for airmass
(quoted at zenith) and are extrapolated - in an overly conservative way - to infrared wavelengths by a factor of the ratio of wavelengths to the $-1/5$ power.

\subsection{Methods}

Data were taken to compare seeing limited imaging with that of tip-tilt only and GLAO corrected imaging. The images were captured with the science 
instrument PISCES, a near-infrared camera with a 110 arc second field of view and a plate scale of 107 milli arc seconds per pixel. Observations comprised
consecutive 1 s exposures, with a rate of approximately 14 exposures per minute. During subsequent analysis, exposures were first background subtracted
and flat fielded, then averaged to simulate a long exposure. Sky de-rotation was then applied post-facto for images with more than one object. All images were
taken with a standard $K_s$ or $\lambda$ = 2.14 $\mu$m narrowband filter. Image widths were calculated using the MOFFAT radial fit tool in the imexam
package of IRAF.

\section{Observations and analysis}
\subsection{February 2008}

The first astronomical targets observed on the night of 2008 February 19 were a series of single stars ranging in visual magnitude from 8 to 10, all with a
declination of approximately +40 degrees. The stars were located approximately in the center of the laser beacon constellation with tip-tilt sensing done
using the target star. Figures 2 and 3 show examples of the stellar PSF during seeing limited, tip-tilt only, and GLAO correction at $K_s$ and $\lambda$ = 2.14 $\mu$m.
Each image is an average of approximately 60 one second exposures.

Star 1 has a seeing limited FWHM of 0.70 arc seconds, while the closed-loop PSF has a FWHM of 0.33 arc seconds,
a reduction of the image width of 53\%, with a factor of 2.3 increase in relative peak intensity. The GLAO corrected PSF shows some asymmetry which
is similar in each of the 60 exposures used to create the mean image in Figure 2 indicating that the system was limited by uncorrected non-common path
errors. The reduction in image width with GLAO can be compared with the cumulative seeing distribution measured at the MMT as presented in Figure 4.
The GLAO correction represents an effective improvement in seeing from somewhat worse than median for the site, at the 58th percentile, to excellent,
at the 4th percentile. 

During observations of star 2, the GLAO correction was modest. A 10\% decrease in image width was observed with tip-tilt only correction; however,
with full ground-layer correction, the FWHM of the image is reduced by 35\%, from 0.60 to 0.39 arc seconds. This corresponds to an effective seeing 
improvement from the 42nd to the 9th percentile.

Although the results from the February 2008 telescope run demonstrate successful closed loop GLAO operation, the level of correction did not achieve 
the final expected system performance of $0.1$ to $0.2$ arc second images in the $K$ band. This is attributable to three factors.  Time to implement and test
the system for measuring and correcting non-common path and other static aberrations was cut short due to inclement weather. This feature was 
implemented and initial testing was completed in the subsequent May 2008 run. Second, and more seriously, there were random network transmission delays
and dropouts in the tip-tilt control loop, which unbeknownst to us at the time, severely affected loop stability. In addition, the elevation axis servo control system of the 
MMT telescope was exhibiting technical problems resulting in the amplification of wind-driven oscillations at 2.25 Hz. The AO loop gain had to be significantly
reduced in the presence of the latter two factors in order to maintain stable closed-loop GLAO operation, compromising the final level of image correction.

\subsection{May 2008}

Additional observations were carried out in 2008 May in order to implement and test automated static aberration correction with the NGS WFS and
to characterize the uniformity of the GLAO corrected PSF over the $110$ arc second field. Unfortunately, these results were again severely affected by the tip-tilt
network issues and continuing problems with the MMT telescope elevation axis servo control system. 
Automated static aberration correction for the MMT LGS WFS system is achieved by measuring the difference in the long term average open-loop
and closed-loop GLAO aberrations of the tip-tilt star with the NGS WFS, similar to the quasi-static calibration with the Keck LGS AO system
\cite{keck3}. The change in the mean NGS WFS slopes is reconstructed to produce a modal estimate of the quasi-static non-atmospheric aberrations.
These modal corrections are converted to LGS WFS centroid offsets using the LGS WFS influence matrix. Finally, these offsets are used to shift the
zero points of the LGS WFS slope calculation during GLAO closed-loop operation. 

Data were taken of an $m_V$ = 9 star ($20^h01^m16.72^s$, $20^\circ41\arcmin29.8\arcsec$, J2000) with many field stars seen in the 110 arc second PISCES field of view. 
Figure 5 shows
an image of the field, with the brighter stars magnified, shown both uncorrected and with ground-layer correction. The stellar images are rescaled in
intensity for clarity. Both corrected and uncorrected data are averages of 70 one second exposures with the narrowband $\lambda$ = 2.14 $\mu$m
filter. Tip-tilt signals were obtained from the bright star in the center of the field, which was also used to sense static aberrations. Image improvement
was again modest: the tip-tilt star has a measured FWHM of 0.70 arc seconds in open-loop and 0.55 arc seconds in GLAO closed-loop, a decrease
of only $21\%$. Over the entire field, both the uncorrected and GLAO corrected PSFs are fairly constant, with a mean seeing limited FWHM 
of $0.72 \pm 0.01$ arc seconds for the brightest 12 stars in this field and a mean GLAO corrected FWHM of $0.58 \pm 0.03$ arc seconds
for the same stars.  In contrast to the results from the February run in Figure 2, the corrected PSFs are round, showing the effect of removing the static
aberration by reference to the NGS WFS. We expect a substantial overall improvement
in the PSF FWHM during future observations now that the vibration issues have been
resolved and we are able to increase the GLAO loop gain, however, a decrease in closed-loop PSF uniformity may be seen due to anisoplanatic effects.

During these observations, the NGS WFS was passively collecting data in both open- and closed-loop operation of the tip-tilt star at 180 frames 
per second. These data have been analyzed to evaluate the closed-loop performance in a manner independent of static aberrations. Table 1 shows
the RMS modal amplitudes reconstructed from the NGS WFS measurements, grouped by radial order.  We show the more familiar Zernike
modes rather than the DH modes used in the real-time control. GLAO corrected $42\%$ of the measured total RMS wavefront error, with
a reduction in the modes controlled by the LGS WFS (orders 2 through 8) of $38\%$, consistent with 
previous open-loop studies of GLAO performance at the MMT\cite{spie}. However, the reduction of $43\%$ for the tip-tilt modes
is much lower than the $>80\%$ correction obtained in previous analyses and attributable to the low tip-tilt loop gains and the
servo oscillations experienced during this run. Note that even though there was a resonance in the telescope's elevation drive, the total power in the
orthogonal tilt modes were quite similar during both open- and closed-loop observations, differing by at most $11\%$, and as consequence would not have affected
an asymmetry in the stellar PSFs. 

\subsection{Control system analysis}

There were two different control loops used in the closed-loop GLAO system, one for the NGS tip-tilt sensor and the other for the 
high-order LGS wavefront sensor. Both loops used an integral control law and had a sample rate of 400 Hz.
The tip-tilt loop was unfortunately hindered by two technical issues. It was discovered during examination of
telemetry data that there were randomly varying network delays and dropouts in the transmission of the tip-tilt data between the tip-tilt sensor and real-time wavefront
reconstruction computer. This would cause the tip-tilt control loop to act erratically and not follow the atmospheric tip-tilt signal.
In addition, the 2.25 Hz tilt signal produced by the oscillating elevation drive increased the required actuator stroke at the edges of the ASM. 
With the combination of these effects, the ASM would quickly hit its safety limits for voice coil actuator current at the edges, breaking the loop - for even modest gains in the tip-tilt loop.
To maintain closed-loop operation for any appreciable
amount of time, the gain was severely reduced, compromising correction. 
Figure 6 shows the open- and closed-loop power spectra for selected Zernike modes as measured by the NGS WFS. The power spectra show that
despite low gain, the GLAO control is removing a significant portion of the aberrations at very low frequencies, below $\sim1$ Hz, for the tip-tilt modes,
and even a $46\%$ reduction in the $2.25$ Hz oscillation in elevation. The random network transmission delays however created a large overshoot in higher
frequencies, hindering the ability to recover good image quality. 

The disturbance rejection frequency response of high-order loop, which was not affected by the tip-tilt issues, was modeled at different gain values (Figure 7). 
At a gain of 0.15, the disturbance rejection
bandwidth is $\sim$ 20 Hz with a low frequency slope of 20 dB/decade. 
To maintain closed-loop operation of the adaptive optics system during observations, the overall system gain was reduced to approximately 0.015. At this gain, the bandwidth
is limited, in the range of $3 - 5$ Hz.
The power spectra of higher order modes (e.g. for the astigmatism modes shown in Figure 6) show correction out to a higher bandwidth of  $\sim2.5$ Hz due to the limited 
influence from the elevation servo resonance. While it is possible that the bandwidth of the higher order modes could have been improved by
taking advantage of the mode dependent gains, a conservative approach to controlling the ASM was adopted, and only the overall system gain was adjusted
to achieve loop stability. With the elevation axis servo oscillation issue
now resolved, we expect the closed-loop bandwidth during future observations to increase to $\sim 20$ Hz once the loop gain is increased to 0.15 and the modal
gains are fine tuned. We intend to also implement a more sophisticated proportional-integral-derivative controller which should further increase the bandwidth of the 
controller to $\sim30$ Hz, significantly reducing residual wavefront error.

\section{Conclusions}

The multiple LGS AO program at the MMT is exploring the practical techniques required for wavefront sensing for ground-layer
and tomographic correction. Closed-loop GLAO operation was successfully demonstrated in February 2008 where image correction reduced stellar
image widths by as much as $53\%$ at $\lambda$ = 2.14 $\mu$m. Wide field,
$\sim2$ arc minute diameter, correction was demonstrated in May 2008 upon successful testing of static aberration
correction and additional system characterization. Optimization of the system 
in subsequent commissioning runs in the fall of 2008 will further improve correction
performance and is expected to deliver $0.1$ to $0.2$ arc second resolution during a majority of seeing conditions. Additional
narrow field PSF characterization will be done in the thermal infrared (3.5 -- 4.8 $\mu$m) where Strehl ratios of $30\%$ to $40\%$ 
are expected using Clio with a plate scale of 48 milli arc seconds per pixels. Imaging with the Clio and PISCES cameras will allow
for the exploration of parameters which affect the ground-layer AO correction. In particular, variables such as control gain, 
reconstruction basis set and the number of controlled modes will be optimized. The effect of observing conditions, such as the 
brightness and field location of the tilt star, will also be explored to support science programs by anticipating the improved resolution
and sensitivity of ground-layer AO correction.

Early shared risk scientific programs will focus on seeing improvement with GLAO, taking advantage of existing near infrared 
instrumentation \citep{mmtsci}, with plans for developing further capability over larger fields of view \cite{loki}. The exploitation of 
routine near infrared seeing of $0.2$ arc seconds or better over a field of 
several arc minutes is likely to be very productive, both for imaging and high resolution multi-object spectroscopy where the 
many-fold improvement in encircled energy within $0.2$ arc seconds will be of particular value.

Development of the MMT laser adaptive optics system will lead to a greater understanding of the important factors in designing
and operating future multiple guide star adaptive optics systems such as the ones being planned for the 8.1-m Gemini North \citep{gemini1}
and Gemini South \citep{gemini2} telescopes, the 8.2-m VLT\citep{ galacsi, vlt, hawki}, the $2 \times 8.4$-m Large Binocular
Telescope\\\citep{lbtao, nirvana}, the 10-m Keck\\\citep{keck, keck2} telescopes, the 25.4-m Giant Magellan Telescope \\\citep{gmt},
30-m Thirty Meter Telescope\citep{tmt}, and 42-m European Extremely Large Telescope\\\citep{eelt}.

\acknowledgments

Observations reported here were made at the MMT, a joint facility of the University of Arizona and the Smithsonian Institution. We are grateful for the continued
support of the MMT Observatory staff, particularly M. Alegria, A. Milone and J. McAfee. We would like to thank T. Pickering for providing the MMT
seeing data. This work has been supported by the National Science Foundation under grants AST-0138347 and AST-0505369 and by the University of
Arizona Technology and Research Initiative Fund. 

{\it Facilities:} \facility{MMT (LGS, PISCES)}





\clearpage

\begin{deluxetable}{lcrrrrrrrrrrrrr}
\tablecolumns{15}
\tabletypesize{\scriptsize}
\tablewidth{0pc}
\tablecaption{Comparison of residual wavefront error with and without GLAO compensation.}
\tablehead{
\colhead{} &  \colhead{}& \multicolumn{2}{c}{Zernike Mode} & \colhead{} &\multicolumn{10}{c}{Zernike Order}\\

\cline{3-4} \cline{6-15}  \\

\colhead{Correction method} & \colhead{Set number}   & \colhead{EL-tilt} & \colhead{AZ-tilt} & \colhead{} & \colhead{1} & \colhead{2} & \colhead{3} &\colhead{4} & \colhead{5} 
& \colhead{6}   & \colhead{7} & \colhead{8} &\colhead{1--8} & \colhead{2--8}    }

\startdata
Uncorrected & 1 & 830 (0.109) & 778 (0.102) &  & 1137 & 486 & 313 & 230 & 187 & 153 & 131 & 118 & 1330 & 690\\
                    & 2 & 868 (0.114) & 867 (0.114) &  & 1227& 460 & 307 & 221 & 175 & 145 & 127 & 114 & 1393 & 660\\
GLAO          & 1 & 477 (0.062) & 473 (0.062) &  & 671 & 244 & 213 & 145 & 130 & 116 & 115 & 107 &  794 & 425\\
                    & 2 & 501 (0.066) & 452 (0.059) &  & 675 & 235 & 209 & 138 & 125 & 109 & 108 & 100 &  790 & 409\\
 \hline          
Mean \% correction & & 42 & 44 & & 43 & 49 & 32 & 37 & 30 & 24 & 14 & 11 &  42 & 38 \\
\enddata

\tablecomments{RMS stellar wavefront error in nm, of individual modes or summed in quadrature over all the
modes in the indicated Zernike orders. Parenthesized values represent the RMS image position error in arc seconds. The bottom
row represents the mean percentage correction with GLAO.}

\end{deluxetable}

\clearpage

\begin{figure}
\begin{center}
\includegraphics[width=4.5in]{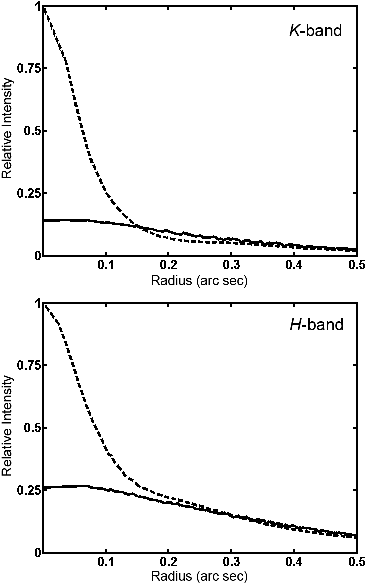}
\end{center}
\caption{Radially averaged synthetic PSF profiles in $K$ and $H$ band calculated from open-loop GLAO correction at the MMT during median seeing\cite{opex}.
The seeing-limited PSFs are shown in solid, while the GLAO-corrected PSFs are shown as dashed lines.}
\end{figure}

\clearpage

\begin{figure}
\plotone{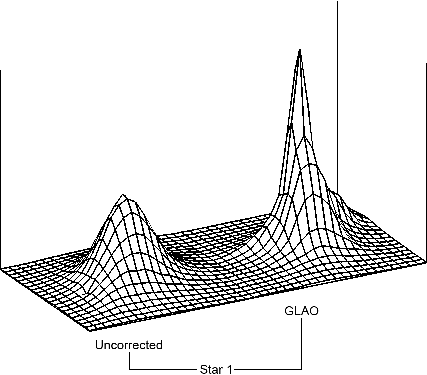}
\caption{Comparison of PSFs for star 1 at $\lambda =$ 2.14 $\mu$m. Without correction, the FWHM of the PSF is 0.70 arc seconds. With GLAO correction,
the FWHM is 0.33 arcseconds. The peak intensity of the GLAO-corrected PSF is 2.3 times greater than the peak intensity of the seeing-limited PSF. Each grid point represents 107 milli arc seconds.}
\end{figure}

\clearpage

\begin{figure}
\plotone{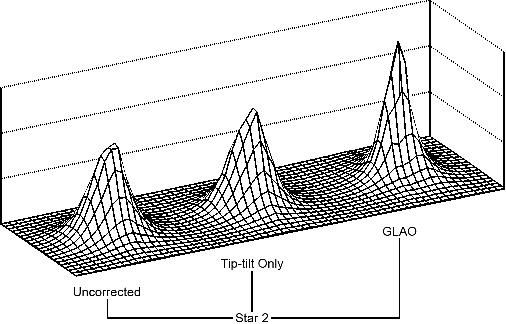}
\caption{Comparison of PSFs for star 2 at $K_s$. Without correction, the FWHM of the PSF is 0.60 arc seconds. With tip-tilt only correction, the FWHM is 0.56 
arc seconds, with an increase in peak intensity of 8\%. With GLAO correction, the FWHM is 0.39 arc seconds, with an increase of peak intensity over the seeing-limited PSF
of 48\%. Each grid point represents 107 milli arc seconds.}
\end{figure}

\clearpage

\begin{figure}
\plotone{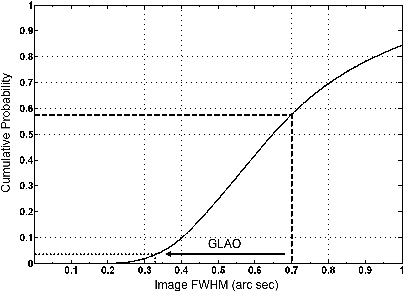}
\caption{A conservative estimate of the MMT's cumulative seeing diagram for $\lambda$ = 2.14 $\mu$m, extrapolated from measurements recorded at visible
wavelengths over a five year period. Observations of star 1 are indicated showing the native seeing (dashed line) and during
GLAO correction (dotted line).}
\end{figure}

\clearpage

\begin{figure}
\plotone{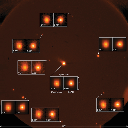}
\caption{PISCES field showing 4 $\times$ magnified images of PSFs with and without ground-layer correction. Each box is linearly
scaled and normalized to the peak intensity while the background image is logarithmically stretched to give greater visibility of faint field stars.
The laser guide stars are located just outside this field on a 120 arc second diameter pentagon.}
\end{figure}

\clearpage

\begin{figure}
\plotone{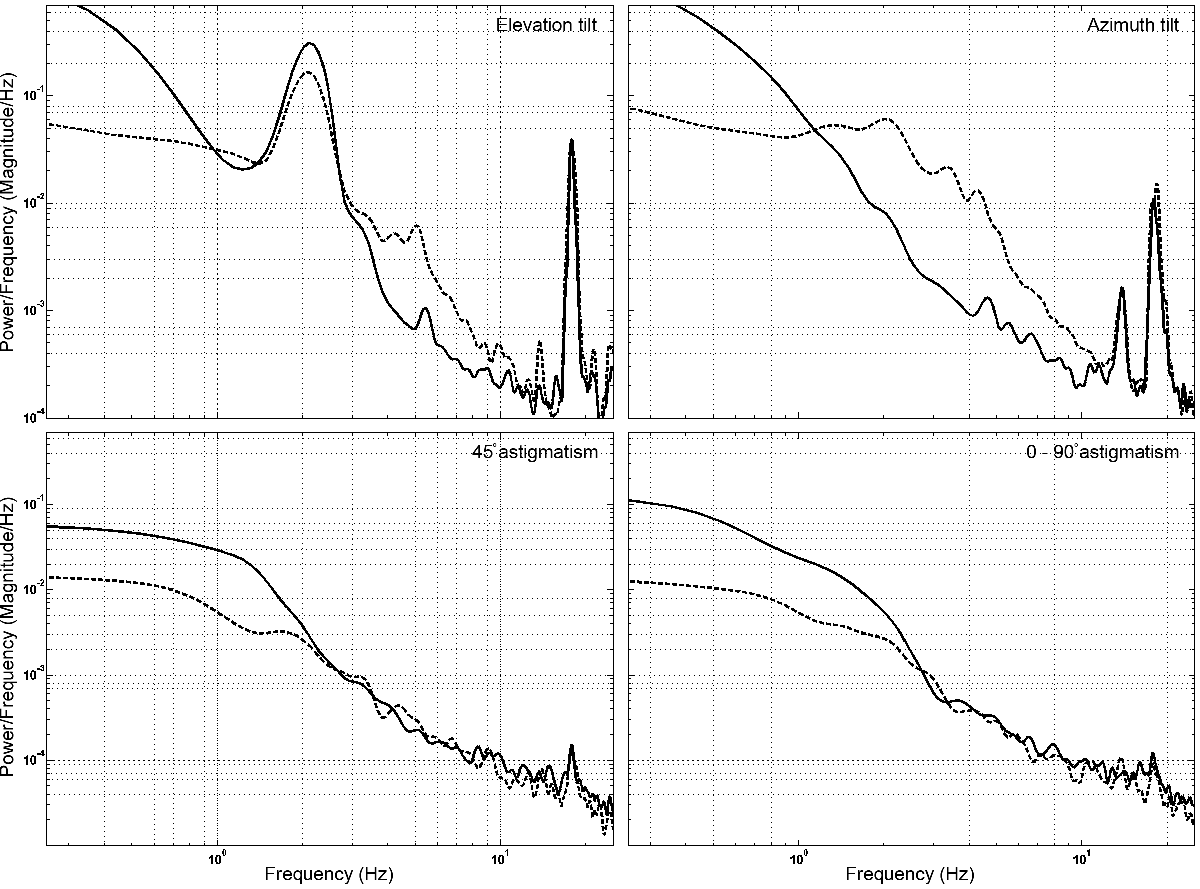} 
\caption{Power spectra of elevation tilt (upper left), azimuth tilt (upper right), $\pm45^\circ$ astigmatism (lower left), and
$0 - 90^\circ$ astigmatism (lower right), during open-loop (solid line) and closed-loop GLAO correction (dashed line).
The peak at 2.25 Hz is due to the elevation servo drive oscillation and the small peak at $\sim18$ Hz is a vibrational
mode of the secondary hub.}
\end{figure}

\clearpage

\begin{figure}
\plotone{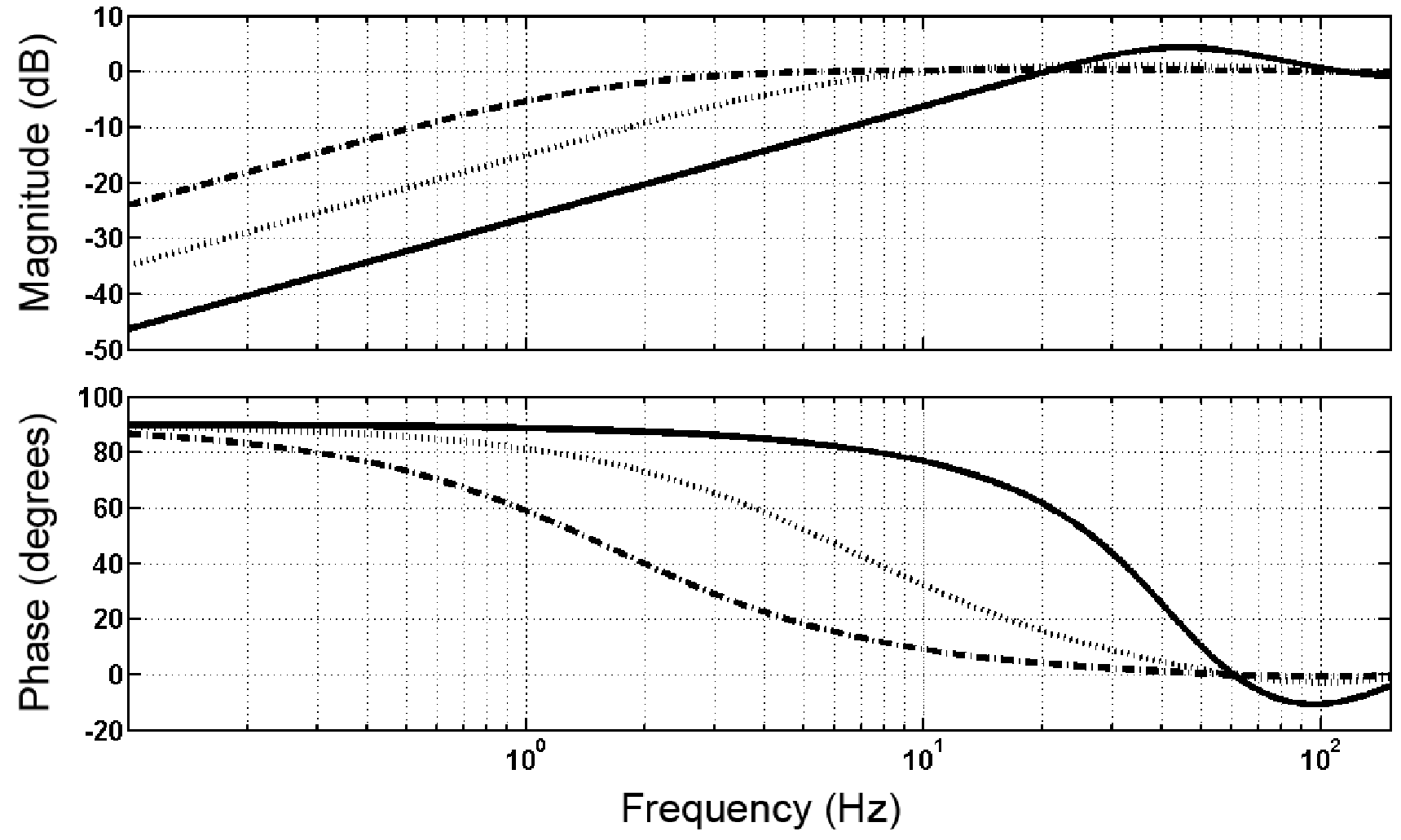} 
\caption{Modeled disturbance rejection frequency response of the high-order modes in the GLAO control system with a sample
rate of 400 Hz. System gains of 0.15 (solid line), 0.05 (dotted line), and 0.015 (dot-dashed line) are
shown.}
\end{figure}

\end{document}